\pdfoutput=1

\documentclass[11pt]{article}

\usepackage[final]{acl}

\usepackage{times}
\usepackage{booktabs}
\usepackage{amsmath}
\usepackage{graphicx}
\usepackage{tabularx}
\usepackage{multirow}
\usepackage{makecell}
\usepackage{array}
\usepackage{listings}
\usepackage{xcolor}
\usepackage{latexsym}

\lstdefinelanguage{json}{
    basicstyle=\small\ttfamily,
    numbers=left,
    numberstyle=\tiny,
    stepnumber=1,
    numbersep=8pt,
    showstringspaces=false,
    breaklines=true,
    frame=lines,
    backgroundcolor=\color{gray!5},
    stringstyle=\color{blue!70!black},
    keywordstyle=\color{purple!70!black},
    commentstyle=\color{gray}
}

\usepackage[T1]{fontenc}

\usepackage[utf8]{inputenc}

\usepackage{microtype}

\usepackage{inconsolata}

\usepackage{graphicx}
\include{latex/math_commands}
\title{DeepXiv-SDK: An Agentic Data Interface for Scientific Literature}

\author{Hongjin Qian, Ziyi Xia, Ze Liu, Jianlyu Chen, Kun Luo, Minghao Qin, Chaofan Li, \\ 
\textbf{Lei Xiong, Junwei Lan, Sen Wang, Zhengyang Liang, Yingxia Shao, Defu Lian, Zheng Liu\thanks{Corresponding author.}}\\
        Beijing Academy of Artificial Intelligence \\
        \texttt{\{chienqhj,zhengliu1026\}@gmail.com} \\
        Project Page: \url{https://github.com/DeepXiv/deepxiv_sdk } 
}

\begin{document}
\maketitle

\begin{abstract}
LLM-agents are increasingly used to accelerate the progress of scientific research. Yet a persistent bottleneck is \emph{data access}: agents not only lack readily available tools for retrieval, but also have to work with unstrcutured, human-centric data on the Internet, such as HTML web-pages and PDF files, leading to excessive token consumption, limit working efficiency, and brittle evidence look-up. This gap motivates the development of \textit{an agentic data interface}, which is designed to enable agents to access and utilize scientific literature in a more effective, efficient, and cost-aware manner. 

In this paper, we introduce \textbf{DeepXiv-SDK}, which offers a three-layer agentic data interface for scientific literature. 1) \textbf{Data Layer}, which transforms unstructured, human-centric data into normalized and structured representations in JSON format, improving data usability and enabling progressive accessibility of the data. 2) \textbf{Service Layer}, which presents readily available tools for data access and ad-hoc retrieval. It also enables a rich form of agent usage, including CLI, MCP, and Python SDK. 3) \textbf{Application Layer}, which creates a built-in agent, packaging basic tools from the service layer to support complex data access demands. 

DeepXiv-SDK currently supports the complete ArXiv corpus, and is synchronized daily to incorporate new releases. It is designed to extend to all common open-access corpora, such as PubMed Central, bioRxiv, medRxiv, and chemRxiv. We release RESTful APIs, an open-source Python SDK, and a web demo showcasing {deep search} and {deep research} workflows. DeepXiv-SDK is free to use with registration.

\end{abstract}

\section{Introduction}
LLM-based agents have emerged as a practical paradigm for turning general-purpose language models into goal-directed systems that can decompose tasks, invoke tools, and refine decisions through iterative feedback~\citep{yao2022react,agentsurvey}. Among their applications, \emph{research agents} are particularly promising for supporting inquiry and evidence-driven scientific decision making~\citep{schmidgall2025agent,Asai2026}. A foundational capability in~such workflows is reliable access to academic papers: agents must quickly identify relevant work, navigate long and heterogeneous documents, and retrieve verifiable evidence to ground claims and synthesis~\citep{mei2025survey,ju2025wispaper,deepresearch}.

Despite rapid progress in agent frameworks, paper access in today’s pipelines remains largely \emph{ad hoc}~\citep{ifargan2025autonomous,miao2025paper2agent}. A common workflow queries a general-purpose search engine, opens a paper in PDF or HTML form, heuristically extracts text, and then feeds large chunks back into the agent for retrieval or question answering~\citep{agarwal2024litllm}. This pipeline is both inefficient and brittle: it repeatedly incurs substantial parsing and reading overhead, depends on document-specific formatting quirks, and lacks a standardized interface across venues and domains~\citep{yang2025agentic}. As a result, intermediate representations are hard to reuse across tasks or agents, and models must reason over noisy, unstructured text without explicit notions of structure, cost, or evidence scope~\citep{qian2025model}.

\begin{figure*}[t]
    \centering
    \includegraphics[width=\linewidth]{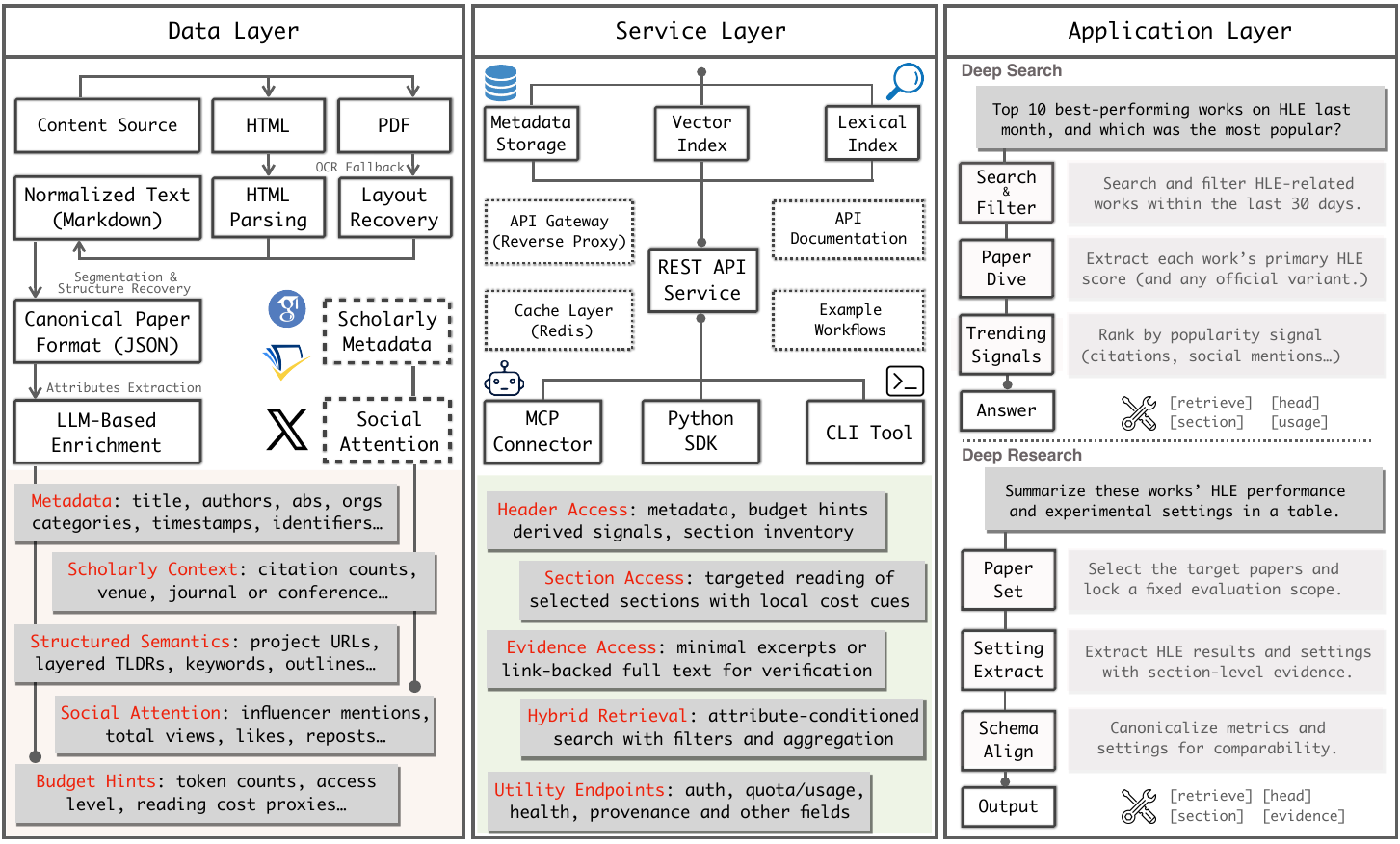}
    \caption{System overview of DeepXiv-SDK. The system ingests and enriches papers into a normalized schema with budget-aware, progressive access views (header-first triage, section-level navigation, and evidence-level verification), and serves them via a REST API backed by hybrid retrieval (lexical and dense indexes). These capabilities support agentic applications including deep search, deep research, and reproducible, evidence-grounded comparison.}
    \label{fig:framework}
    \vspace{-10pt}
\end{figure*}

An agent-friendly solution should treat paper access as a \emph{data interface}, not a one-off parsing step~\citep{yang2025agentic}. First, it should be \textbf{structured and normalized}, exposing papers through a consistent schema so agents can access metadata, document structure, and supporting evidence via a single protocol~\citep{tirado2016web,lu2025build}. Second, it should support \textbf{progressive disclosure by knowledge density}, offering coarse-to-fine views that allow agents to decide \emph{what to read} and \emph{how much to read} before paying the full-context cost, thereby reducing cognitive burden and unnecessary token expenditure~\citep{qian2025model}. Third, it should be \textbf{retrieval-oriented and conditionable}, enabling agents to locate and curate papers by composing constraints over multiple attributes and then routing to the most relevant parts once candidates are identified~\citep{song2025beyond}. Together, these principles make paper access reusable, budget-aware, and evidence-seeking.

Guided by these principles, we introduce \textbf{DeepXiv-SDK}, a unified, agent-callable interface that turns papers into \emph{structured objects with controllable access cost}. DeepXiv-SDK materializes each paper into schema-normalized views that an agent can query deterministically: a header-first view that exposes core metadata, section inventory, and global budget cues; a section-addressable view that supports targeted reading without full-document ingestion; and an evidence-level view that returns full content for verification and downstream processing. 
To make discovery and curation equally tool-friendly, DeepXiv-SDK also provides hybrid, attribute-conditioned retrieval over lexical and dense indexes, allowing agents to filter and aggregate candidates by practical constraints (e.g., category signals, authors, time ranges, citation/venue attributes) before drilling into the most relevant sections. We deploy DeepXiv-SDK at \textbf{arXiv scale} with \textbf{daily synchronization} to new releases (typically within 24 hours) and release RESTful APIs, an open-source SDK, and a live demo showcasing deep-search and deep-research workflows.

Finally, we construct a small \textbf{practical evaluation set} to assess end-to-end usefulness under two task families that match the interface design. In \emph{deep search}, agents retrieve and shortlist relevant papers under time constraints using retrieval plus header-first screening; in \emph{deep research}, agents selectively read sections to extract detailed evidence, and produce evidence-linked reports or comparison tables. Across both settings, DeepXiv-SDK reduces token overhead by avoiding default full-text ingestion, improves retrieval precision via hybrid, attribute-conditioned search and structure-aware routing, and yields higher-quality outputs by escalating to evidence-level access only when verification is needed. We also stress-test the service and observe that the current stack sustains \emph{multi-million requests per day} with scale-out capability, supporting interactive agent use in practice.

\section{Related Work}
\label{sec:related}
As LLM agents increasingly rely on tool use and multi-step interaction, \emph{data access} often becomes a bottleneck for scaling agent capability, since agents must repeatedly retrieve, parse, and ground on external artifacts such as web pages and PDFs~\citep{agentsurvey,yao2022react,deepresearch}. A growing line of work therefore argues for turning raw web artifacts into \emph{structured, agent-callable interfaces} with normalized schemas, controllable views, and provenance to reduce ad hoc parsing and improve reliability~\citep{qian2025model,song2025beyond}. 
Within scientific literature, many prior efforts focus on agent frameworks or web platforms to improve literature review efficiency (e.g., search-then-read workflows, iterative summarization, and multi-document QA), but typically do not expose a reusable \emph{data interface API} that agents can call deterministically across tasks~\citep{pasa,miao2025paper2agent,ju2025wispaper}. 
The closest efforts in spirit are ar5iv and AlphaXiv, which provide more usable browsing experiences by converting arXiv papers into readable HTML and enriched views.\footnote{\url{https://ar5iv.labs.arxiv.org/}}\footnote{\url{https://www.alphaxiv.org/}} These systems are best viewed as advanced mirrors of arXiv for human browsing, rather than agentic data interfaces designed to scale through a reusable API and progressive, budget-aware access. 

In contrast, DeepXiv-SDK provides an accessible, progressive interface with explicit budget cues and hybrid, attribute-conditioned retrieval, enabling agents to screen cheaply, read selectively, and verify on demand.

\section{System: DeepXiv-SDK}
\subsection{Overview}
\label{subsec:overview}
Figure~\ref{fig:framework} presents \textbf{DeepXiv-SDK} as a paper-native, agentic data interface that turns academic papers into \emph{structured, tool-callable objects} with controllable access cost. The system contains three layers.

The \textbf{Data Layer} performs corpus-scale normalization and enrichment, materializing a \emph{section-addressable} canonical representation with machine-consumable signals such as document structure, lightweight summaries, and explicit budget cues (e.g., token/length statistics). Building on this, the \textbf{Service Layer} exposes a unified protocol that provides (i) \emph{progressive access} via structured views that increase in information density and cost (header, section, evidence), and (ii) \emph{hybrid, attribute-conditioned retrieval} for constructing and refining candidate paper sets before reading. Finally, the \textbf{Application Layer} instantiates these primitives into end-to-end demo services, including \emph{deep search} for candidate discovery and screening and \emph{deep research} for iterative section reading and evidence-linked synthesis. This layered design keeps the interface modular and reusable; we detail each layer in the following sections.
\subsection{Data Layer: Corpus-Scale Ingestion, Structuring, and Signal Materialization}
\label{subsec:data-layer}
The \textbf{data layer} (Figure~\ref{fig:framework}, left) converts heterogeneous arXiv artifacts into \emph{schema-stable, agent-consumable paper objects}. arXiv provides official metadata, while content is delivered as PDF/HTML with highly variable layout and section cues; directly reading these raw artifacts forces every agent pipeline to re-implement parsing and segmentation, leading to brittle failures and non-reproducible reads. DeepXiv-SDK centralizes this work and materializes a canonical representation with explicit structure, derived signals, and budget cues.

\begin{table}[t]
\centering
\small
\setlength{\tabcolsep}{4pt}
\renewcommand{\arraystretch}{1.12}
\begin{tabular}{@{}p{0.30\linewidth} p{0.66\linewidth}@{}}
\toprule
\textbf{Signal / View} & \textbf{Materialized content (examples)} \\
\midrule
\textbf{Core metadata} &
title, authors, abstract, categories, publish/update time, affiliations, identifiers (arXiv ID, DOI if available) \\

\textbf{Structured info} &
section outline, section TL;DRs, keywords, resource links \\

\textbf{Budget hints} &
token/length estimates (paper- and section-level), preview truncation flags (e.g., \texttt{is\_truncated}, total characters) \\

\textbf{Scholarly context} &
citation attributes and venue signals when available (stored with provenance) \\

\textbf{Social attention} &
optional attention indicators aggregated from posts linking \texttt{arxiv.org} (e.g., views, likes, reposts) \\
\addlinespace[2pt]
\midrule
\textbf{Overview view} &
header-first payload: core metadata, derived signals, section inventory, global budget hints \\

\textbf{Section view} &
section-addressable payloads: section text, summaries, per-section budget hints \\

\textbf{Evidence view} &
verification-ready full content: full Markdown and structured JSON for deterministic downstream processing \\
\bottomrule
\end{tabular}
\caption{Data-layer deliverables in DeepXiv-SDK: enriched signal families and materialized access views.  }
\label{tab:data-layer-signals}
\end{table}

\paragraph{Processing pipeline.}
Given an arXiv ID, the data layer runs a deterministic pipeline that produces a section-addressable paper object plus derived signals. It first pulls official metadata via OAI-PMH and records publish/update timestamps for synchronization. It then acquires the source artifact, preferring the HTML-rendered view when available and falling back to the PDF otherwise. For PDFs, we convert the document to Markdown with MinerU~\cite{wang2024mineruopensourcesolutionprecise} to normalize heterogeneous layouts into a text-centric format; for HTML, we extract the main content and normalize it into the same internal representation. Next, we recover document structure by detecting heading cues and formatting regularities, constructing an ordered section inventory (section titles and hierarchy) and segmenting normalized text into section-level payloads. These outputs are assembled into a \textbf{canonical JSON} with a fixed schema (paper-level fields plus an explicit section map), which supports deterministic section addressing across papers.

On top of the canonical JSON, we materialize the signals required for progressive access and retrieval. We first compute \textbf{budget hints}, including global and per-section token/length statistics (via \texttt{tiktoken}). We then generate \textbf{lightweight semantic signals}, including a paper preview TL;DR and per-section TL;DRs, using a small instruction model. Next, we extract \textbf{resource links} (e.g., GitHub repositories) via regex and use the LLM to validate whether each candidate URL is truly associated with the paper by jointly considering the URL identity and the paper preview context. We further attach optional \textbf{external context} when available, including citation counts and venue/journal metadata by linking arXiv IDs to third-party scholarly services such as Semantic Scholar or Google Scholar, as well as \textbf{social attention} signals aggregated from X.com search results for \texttt{arxiv.org} mentions (e.g., views, likes, reposts). Finally, we persist multiple \textbf{materialized views} (overview, section, and evidence forms) together with provenance fields (source type, extraction time, update time), enabling traceable and reproducible access in downstream agent pipelines.

\paragraph{Delivered products.}
The data layer outputs a bundle of materialized views together with enriched signal families, so agents can consume papers deterministically without ad hoc parsing. Concretely, DeepXiv-SDK provides an \textbf{overview view} for screening and routing, a \textbf{section view} for section-addressable navigation, and an \textbf{evidence view} that exposes full content. These views are designed to support an explicit escalation path from low-cost triage to targeted reading and, when needed, evidence-level verification. Across all views, responses include budget hints (paper- and section-level cost cues) and provenance (source type, extraction time, update time), enabling agents to plan reading cost, track freshness, and retain traceability. Table~\ref{tab:data-layer-signals} summarizes the materialized views and the corresponding signal families.

\subsection{Service Layer: Unified Protocol for Progressive Access and Hybrid Retrieval}
\label{subsec:service-layer}
The \textbf{service layer} turns data-layer artifacts into a \emph{stable, budget-aware} interface that agents can invoke. The goal is to make paper access (i)~\textbf{protocol-reusable} across tasks and corpora, and (ii)~\textbf{cost-controllable} so agents can start with low-cost screening, escalate to section-level reads, and only request full text when verification is required. DeepXiv-SDK exposes these primitives through a unified REST service with bearer-token authentication, Redis caching for high-frequency reads, on-demand loading for heavier views, and a usage endpoint that makes tool calls auditable.

\paragraph{API surface.}
Table~\ref{tab:service-layer-api} summarizes the core endpoints. The API provides two complementary capabilities. First, it supports \textbf{progressive access} via structured views that increase in information density and cost (overview, preview, section, full text). Second, it provides \textbf{hybrid retrieval} (lexical plus dense) with attribute conditioning, enabling agents to build and refine candidate paper sets before reading. Together, these endpoints allow agents to control reading budget explicitly and defer evidence-level access until necessary.

\paragraph{Client interfaces as service surfaces.}
To reduce integration friction, DeepXiv-SDK provides three thin clients that bind to the same REST protocol: (i) a \textbf{Python SDK} with deterministic calls for retrieval and progressive reading, (ii) an \textbf{MCP connector} that registers endpoints as tool primitives in agent runtimes, and (iii) a \textbf{CLI} for scripted use and reproducible evaluation. Treating these clients as part of the service layer ensures that the protocol is operationally usable, not merely specified.

\subsection{Application Layer: SDK, Agent, and Deep-Research Workflows}
\label{subsec:application-layer}
The \textbf{application layer} packages the service primitives into \emph{developer- and agent-ready} tooling for academic-paper deep research. The goal is twofold: to provide reproducible bindings that integrate cleanly into research-agent stacks, and to ship an agent implementation that directly instantiate progressive paper access as an executable workflow.

DeepXiv-SDK includes a lightweight Python SDK that wraps the REST protocol into a small set of deterministic calls spanning \emph{retrieval} and \emph{progressive reading} (header, section, and evidence access). On top of these tools, DeepXiv-SDK integrates a \textbf{built-in agent} specialized for paper-centric deep research: users can instantiate it in Python (e.g., \texttt{agent = deepxiv.agent(...)}; \texttt{agent.query("...")}) or invoke it from the CLI (e.g., \texttt{deepxiv agent query "..."}), where the agent automatically retrieves candidates, screens them via low-cost views, routes to relevant sections, and escalates to evidence-level reads only when verification is needed.

To make the intended behavior concrete, we expose two canonical workflows. \emph{Deep search} emphasizes candidate set construction, filtering, and ranking via hybrid retrieval and header-level signals (optionally incorporating social attention indicators). \emph{Deep research} performs iterative section reading to extract experimental settings and results, and produces evidence-linked summaries or comparison tables across a paper set. Finally, we serve a \textbf{live demo} at \href{https://1stauthor.com/}{\textit{this website}} that showcases these workflows end-to-end in an interactive setting.

\section{License}
\label{sec:license}

\textbf{DeepXiv-SDK} is designed as an \emph{agentic access and structuring interface} rather than a full-text redistribution service. We reuse arXiv \emph{descriptive metadata} (e.g., title, authors, abstract, categories, and timestamps) under the arXiv API Terms of Use, which state that arXiv e-print metadata are available under \textbf{CC0 1.0}.\footnote{\url{https://info.arxiv.org/help/api/tou.html}} Beyond metadata, DeepXiv-SDK derives navigation and enrichment signals (e.g., section structure, summaries, budget cues, and optional scholarly/social context) to support agentic screening, routing, and evidence-seeking.

For paper content, we do not claim redistribution rights and do not mirror or bulk-serve full papers. Most arXiv submissions use the default arXiv license, which grants arXiv the right to distribute the work but does not generally grant third parties an unrestricted right to redistribute full text; arXiv further recommends that tools built on full text \textbf{link back to arXiv} for downloads\footnote{\url{https://info.arxiv.org/help/bulk_data_s3.html}}. Accordingly, for any request that requires accessing the original paper, DeepXiv-SDK returns the \textbf{source link} (the arXiv landing page or the publisher-provided URL) rather than serving a mirrored copy. This stance is consistent with existing arXiv-derived browsing interfaces such as ar5iv\footnote{\url{https://ar5iv.labs.arxiv.org/}} and AlphaXiv\footnote{\url{https://www.alphaxiv.org/}}, which improve usability while linking back to the source.

\section{Evaluation}
\label{sec:eval}

We evaluate DeepXiv-SDK on two tasks that match our interface design: (i) \emph{agentic paper search} under multi-constraint queries, and (ii) \emph{deep research} for complex, evidence-backed QA. We also benchmark service latency and caching. Figure~\ref{fig:performance} summarizes results, and Table~\ref{tab:example-eval-items} gives representative queries.

\subsection{Agentic Search}
\label{subsec:eval-search}

\begin{figure}
    \centering
    \includegraphics[width=\linewidth]{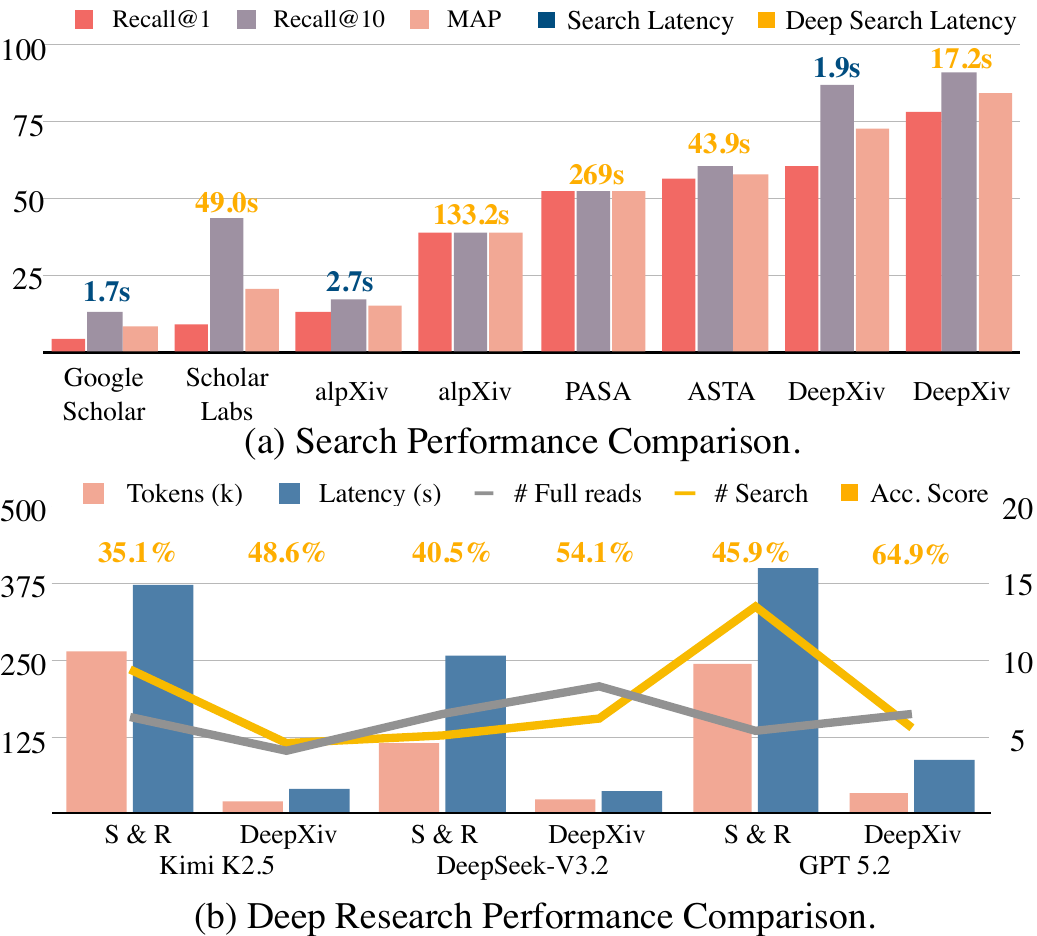}
    \caption{Evaluation of DeepXiv-SDK. (a) Agentic paper search on 50 multi-constraint queries with unique targets: DeepXiv achieves higher Recall@1/10 with substantially lower latency than existing agentic search platforms. (b) Deep research QA on 47 queries: DeepXiv reduces token and time cost while improving answer quality compared to a traditional Search\&Read pipeline.}
    \label{fig:performance}
\end{figure}

\paragraph{Setup.}
We construct a search-focused evaluation set of 50 multi-condition queries, each mapped to a \emph{single} target paper (unique answer). We compare DeepXiv against five academic search platforms that expose agentic search capabilities: Google Scholar, Google Scholar Labs~\footnote{\url{https://scholar.google.com/scholar_labs/}}, alpXiv~\footnote{\url{https://www.alphaxiv.org/}}, PASA~\footnote{\url{https://pasa-agent.ai/}}, and ASTA~\footnote{\url{https://asta.allen.ai/}}. We report retrieval metrics and measure end-to-end latency for plain search versus agentic deep search (Figure~\ref{fig:performance} a).

\paragraph{Results and analysis.}
DeepXiv consistently outperforms the baselines in both accuracy and efficiency. Across the multi-constraint queries, DeepXiv yields stronger top-rank retrieval and maintains high recall while remaining near-interactive for standard search and substantially faster for agentic search than verification-heavy systems. Qualitatively, PASA and alpXiv often enumerate candidates and validate by reading full paper text, whereas DeepXiv leverages \emph{progressive access}: agents first screen with low-cost header/preview signals and only escalate to section- or evidence-level reads when necessary, which improves precision while reducing latency.

\subsection{Deep Research QA}
\label{subsec:eval-deepresearch}

\paragraph{Setup.}
We build a second evaluation set of 47 complex QA queries that require aggregating and verifying paper-derived evidence (e.g., time-scoped questions over benchmark results such as ``best score in the last month''). We compare two agent pipelines: (i) a traditional \emph{Search \& Read} (S\&R) pipeline using Google Search for retrieval and Jina Reader~\footnote{\url{https://jina.ai/}} for reading full text, and (ii) a DeepXiv-based pipeline that uses attribute-conditioned retrieval plus progressive access (header, section, evidence) for targeted reading.

\paragraph{Results and analysis.}
Figure~\ref{fig:performance}b shows that DeepXiv-based deep research reduces both tool cost and wall-clock latency by avoiding default full-text ingestion and replacing brittle page parsing with structured, section-addressable access. Across representative models, DeepXiv improves end-to-end answer quality consistently while substantially reducing token consumption and runtime. This pattern aligns with the intended interaction model: the agent screens candidates using low-cost views, selectively reads relevant sections, and escalates to evidence-level access only when verification is needed. Overall, the results support the central claim of DeepXiv-SDK: making papers \emph{tool-callable} with progressive, budget-aware access improves both efficiency and grounding quality for research-agent workflows.

\subsection{Latency Performance}
\label{subsec:latency}

We benchmark DeepXiv-SDK latency on 1{,}000 arXiv IDs with concurrency $=16$, reporting mean cold (cache-miss) and warm (cache-hit) times for \textbf{Local} (same-region) and \textbf{Remote} (cross-country) calls~(Table~\ref{tab:latency}). The service remains interactive under this load, and caching provides reliable acceleration for frequently accessed views (up to $3.36\times$ on \texttt{preview}); even remotely, warm latencies stay within a few hundred milliseconds for heavier endpoints such as \texttt{json} (181.6\,ms).

For context, a conventional ``fetch+parse'' workflow on the same 1{,}000 papers takes 31m12s to fetch HTML/PDF via a commercial proxy pool and 88m49s to convert to Markdown (with 8$\times$A100 GPUs), totaling 120m01s, or 7.20s per paper. Compared to this baseline, DeepXiv-SDK delivers large end-to-end speedups, e.g., $54.6\times$ (Local) and $39.6\times$ (Remote) for warm \texttt{json} access.

\begin{table}[t]
\centering
\small
\setlength{\tabcolsep}{5pt}
\renewcommand{\arraystretch}{1.10}
\begin{tabular}{p{0.16\linewidth}@{} cccc c}
\toprule
 & \multicolumn{2}{c}{\textbf{Local (ms)}} & \multicolumn{2}{c}{\textbf{Remote (ms)}} & \textbf{Cache spd} \\
\cmidrule(lr){2-3}\cmidrule(lr){4-5}\cmidrule(lr){6-6}
& \textbf{Cold} & \textbf{Warm} & \textbf{Cold} & \textbf{Warm} & \textbf{ (L / R)}\\
\midrule
\texttt{head}    & 23.21  & 12.51  & 142.55 & 114.62 & 1.86 / 1.24 \\
\texttt{brief}   & 187.55 & 164.46 & 307.39 & 196.25 & 1.14 / 1.57 \\
\texttt{raw}     & 108.14 & 146.15 & 216.29 & 167.89 & 0.74 / 1.29 \\
\texttt{json}    & 203.79 & 131.89 & 283.21 & 181.63 & 1.55 / 1.56 \\
\texttt{preview} & 102.05 & 30.39  & 201.46 & 112.31 & 3.36 / 1.79 \\
\bottomrule
\end{tabular}
\caption{Mean latency over 1{,}000 arXiv IDs at concurrency $=16$ (Local vs Remote; cold vs warm). Cache spd is cold/warm. In comparison, fetch+parse baseline: 120m01s for 1{,}000 papers (7{,}200\,ms per paper).}
\label{tab:latency}
\end{table}
\section{Conclusion}
In this paper, we focus on \emph{paper access} as a practical bottleneck for research agents: existing pipelines often treat papers as raw PDF/HTML artifacts, leading to expensive full-text ingestion and brittle evidence lookup. We propose \textbf{DeepXiv-SDK}, an agentic data interface that represents papers as normalized, tool-callable objects with progressive access (header $\rightarrow$ section $\rightarrow$ evidence) and hybrid, attribute-conditioned retrieval. Through an arXiv-scale deployment with daily synchronization, we operationalize this design as a stable service and SDK that agents can invoke repeatedly across tasks. Empirically, via task-driven evaluation on multi-constraint agentic search and complex deep-research QA, as well as latency benchmarking under concurrent load, we show that DeepXiv-SDK enables cheaper screening, more selective section reading, and verification only when needed, improving both efficiency and evidence grounding in AI4Science workflows.

Implementation details of DeepXiv-SDK (data processing and service deployment) are provided in Appendix~\ref{sec:impl}, current corpus statistics are summarized in Appendix~\ref{sec:stat}, and Listing~\ref{tab:memorag-json} showcases the metadata format returned for an example paper.

\clearpage

\bibliography{custom}

\clearpage
\appendix

\section{Implementation and Deployment}
\label{sec:impl}

DeepXiv-SDK consists of an arXiv-scale processing pipeline that normalizes papers into section-addressable objects with enriched signals, and a serving stack that exposes progressive access and hybrid retrieval through a cached REST interface.

\subsection{ArXiv-Scale Processing Pipeline}
We obtain the full arXiv metadata stream via the \textbf{OAI-PMH} interface, keyed by arXiv ID. For the initial full crawl, we download PDFs for all papers and convert them to Markdown using \textbf{MinerU}~\cite{wang2024mineruopensourcesolutionprecise}, which normalizes heterogeneous layouts into a text-centric representation. This PDF$\rightarrow$MD stage runs on 8 nodes with 8$\times$H100 GPUs each and takes $\sim$72 hours. From Markdown, we recover section structure using formatting regularities and materialize a canonical JSON per paper, where each section is an explicit field to support deterministic section-level access.

On top of the structured JSON, we run lightweight LLM-based enrichment with \textbf{Qwen3-4B-Instruct-2507}. Given the first 2048 tokens and the metadata author list, the model infers author--affiliation relations, produces a one-sentence TL;DR, and extracts up to five keywords. We further extract GitHub repository links with regex and filter common non-paper-specific repositories (e.g., \texttt{pytorch}, \texttt{vllm}, \texttt{sklearn}); additionally, the model validates whether each extracted GitHub URL is \emph{paper-specific} by cross-checking the repository identity (e.g., owner/name) against the paper context. For budget-aware access, we compute total and per-section token counts using \textbf{tiktoken} and generate a TL;DR for each section. The enrichment stage runs on the same 8$\times$(8$\times$H100) cluster and takes $\sim$100 hours for the full corpus.

\begin{table}[t]
\centering
\small
\setlength{\tabcolsep}{6pt}
\renewcommand{\arraystretch}{1.12}
\begin{tabular}{@{}p{0.7\linewidth} p{0.21\linewidth}@{}}
\toprule
\textbf{Question} & \textbf{Answer (arXiv ID)} \\
\midrule
Which paper proposes an imagination-driven agent that balances rescuing others and minimizing environmental harm in conflicting grid-world tasks, outperforming both standard and empathy-based baselines across diverse scenario variants? & 2501.00320 \\
\midrule
Which paper builds a course-specific chatbot using course documents to guide a conversational model, outperforming a general model on database questions and policies, citing sources, with no instructor training? & 2401.00052 \\
\midrule
Which paper develops a general algorithm for constructing minimal diagonally concave functions on strips with arbitrary boundary data, including cases with horizontal herringbone foliations and fissures? & 2401.00053 \\
\bottomrule
\end{tabular}
\caption{Example evaluation items: question-to-paper identification with arXiv IDs as gold answers.}
\label{tab:example-eval-items}
\vspace{-10pt}
\end{table}

\subsection{Serving, Indexing, and \textbf{T+0} Updates}
Processed JSON and Markdown are stored in object storage, while metadata and extracted attributes are stored in \textbf{PostgreSQL}. We build \textbf{Elasticsearch} indexes over attributes and content surrogates, and compute dense embeddings with \textbf{BGE-m3} on a compact representation (title + abstract + section TL;DRs) to support hybrid retrieval without indexing full text. The service is deployed on a 64-core/256GB node (with a hot-standby replica), using \textbf{Gunicorn} for the API, \textbf{Caddy} for reverse proxying and load balancing, and \textbf{Redis} for caching high-frequency endpoints.

For incremental maintenance, DeepXiv-SDK performs weekday scheduled syncs (Mon--Fri) by pulling OAI-PMH listing deltas and processing new/updated IDs. Since many arXiv entries provide HTML pages, we adopt an \textbf{HTML-first} strategy for parsing, falling back to PDF when HTML parsing fails; the downstream structuring and enrichment steps remain identical. This yields \textbf{T+0} synchronization with new releases and keeps retrieval and progressive-access views fresh. Finally, the pipeline is designed to extend to other open-access corpora (e.g., \textbf{PMC}) by swapping only the ingestion connector, while reusing the same normalization schema, enrichment, indexing, and serving protocol.

\section{Statistics}
\label{sec:stat}
As of now, DeepXiv-SDK indexes \textbf{2,949,129} arXiv papers; among them, \textbf{2,712,378} are successfully parsed into an explicit section structure with per-section TL;DRs, while the remaining papers typically lack usable section cues (no section markers, weak/ambiguous structure, or content dominated by figures), yet still support the other API views and retrieval features. Across the corpus, we extract \textbf{219,717} GitHub URLs; a manual audit of 200 sampled links found \textbf{no mismatches}. For social attention, we query the X.com search API daily for \texttt{arxiv.org} mentions and aggregate exposure signals, yielding trending data for \textbf{448,825} papers to date. Finally, we enrich papers with \textbf{citation counts} and \textbf{venue/journal} metadata by linking arXiv IDs to Semantic Scholar records.

\begin{table*}[t]
\centering
\small
\setlength{\tabcolsep}{6pt}
\renewcommand{\arraystretch}{1.10}
\begin{tabularx}{\textwidth}{@{}>{\raggedright\arraybackslash}p{0.24\textwidth}
                             >{\raggedright\arraybackslash}p{0.30\textwidth}
                             >{\raggedright\arraybackslash}X@{}}
\toprule
\textbf{Capability} & \textbf{Endpoint (template)} & \textbf{Returns (role)} \\
\midrule
\multicolumn{3}{@{}l}{\textbf{Progressive access (arXiv)}} \\
\midrule
Quick Brief &
\texttt{/arxiv (type=brief, id=\{id\})} &
Brief metadata. \\

Overview (header-first) &
\texttt{/arxiv (type=head, id=\{id\})} &
Metadata + section inventory + global budget hints. \\

Preview &
\texttt{/arxiv (type=preview, id=\{id\})} &
Fixed-length prefix (10k chars) + truncation flags. \\

Section read &
\texttt{/arxiv (type=section, id=\{id\}, section=\{sec\})} &
Named section payload + local budget cues. \\

Full text (Markdown) &
\texttt{/arxiv (type=raw, id=\{id\})} &
Full Markdown for verification. \\

Full object (JSON) &
\texttt{/arxiv (type=json, id=\{id\})} &
Full structured JSON for deterministic processing. \\

\addlinespace[2pt]
\midrule
\multicolumn{3}{@{}l}{\textbf{Retrieval and utilities}} \\
\midrule
Hybrid retrieval \& filtering &
\texttt{/arxiv (type=retrieve, q=\{\dots\})} &
BM25 / vector / hybrid search with attribute filters and pagination. \\

Social attention (X) &
\texttt{/arxiv/trending\_signal (id=\{id\})} &
Optional exposure stats: total\_views, total\_likes, total\_reposts. \\

Usage \& quota audit &
\texttt{/stats/usage (days=\{n\})} &
Usage summary for quota-aware tool planning. \\

\addlinespace[2pt]
\midrule
\multicolumn{3}{@{}l}{\textbf{Corpus extensibility (PMC)}} \\
\midrule
PMC basic access (current) &
\texttt{/pmc (type=head, id=\{id\})}, \texttt{/pmc (type=json, id=\{id\})} &
Currently serves basic metainfo and full JSON; section parsing/TL;DRs are not yet materialized. \\
\bottomrule
\end{tabularx}
\caption{DeepXiv-SDK API surface. Progressive access controls cost, while hybrid retrieval curates sets before evidence-level inspection. Table~\ref{tab:memorag-json} showcases a header response. We detail documentation at \href{https://data.rag.ac.cn/api/docs}{\textit{this page}}.}
\vspace{-10pt}
\label{tab:service-layer-api}
\end{table*}

\begin{table*}[t]

\centering
\begin{minipage}{0.95\textwidth} 
\begin{lstlisting}[language=json, caption={Example metadata (JSON), accessible at \url{https://data.rag.ac.cn/arxiv/?arxiv_id=2409.05591&type=head}. This ID is token-free; other requests require an API token.}, label={tab:memorag-json}]
{
  "arxiv_id": "2409.05591",
  "src_url": "https://arxiv.org/pdf/2409.05591",
  "title": "MemoRAG: Boosting Long Context Processing with Global Memory-Enhanced Retrieval Augmentation",
  "abstract": "Processing long contexts presents a significant challenge for large language models (LLMs)...",
  "authors": [
    {
      "name": "Hongjin Qian",
      "orgs": [
        "Beijing Academy of Artificial Intelligence", ]}...],
  "token_count": 23311,
  "venue": "The Web Conference",
  "journal_name": "Proceedings of the ACM on Web Conference 2025",
  "citations": 63,
  "sections": [
    {
      "name": "1. Introduction",
      "idx": 0,
      "tldr": "MemoRAG introduces a novel dual-system framework that mimics human cognition by first building a global memory of long documents...",
      "token_count": 1661 }... ],
  "categories": ["cs.CL", "cs.AI"],
  "publish_at": "2024-09-09T00:00:00",
  "keywords": [
    "global memory augmentation", 
    "long-context retrieval", ...
  ],
  "tldr": "[research paper] MemoRAG introduces a dual-system RAG framework that mimics human cognition by first constructing a global memory of long...",
  "github_url": "https://github.com/qhjqhj00/MemoRAG"
}

\end{lstlisting}

\end{minipage}

\end{table*}
\end{document}